\documentclass[preprint2]{aastex}

\shorttitle{Maximum young cluster mass in M~51}
\shortauthors{Gonz\'alez-L\'opezlira et al.}

\begin{document}

\title{Gas surface density, star formation rate surface density, and the maximum mass                   
of young star clusters in a disk galaxy. II. The grand-design galaxy M~51}

\author{Rosa A.\ Gonz\'alez-L\'opezlira\altaffilmark{1,2}, 
Jan Pflamm-Altenburg\altaffilmark{1}, \& Pavel Kroupa\altaffilmark{1}}

\altaffiltext{1}{Argelander Institut f\"ur Astronomie, Universit\"at Bonn, Auf dem H\"ugel 71, D-53121 Bonn, Germany}
\altaffiltext{2}{On sabbatical leave from the Centro de Radioastronom\'{\i}a y Astrof\'{\i}sica, UNAM, Campus Morelia, Michoac\'an, M\'exico, C.P. 58089; 
{\tt r.gonzalez@crya.unam.mx}}
\begin{abstract}
We analyze the relationship between maximum cluster mass, and surface
densities of total gas ($\Sigma_{\rm gas}$), molecular gas
($\Sigma_{\rm H_2}$), neutral gas ($\Sigma_{\rm HI}$) and star
formation rate ($\Sigma_{\rm SFR}$) in the grand design galaxy M~51,
using published gas data and a catalog of masses, ages, and reddenings
of more than 1800 star clusters in its disk, of which 
223 are above the cluster mass distribution function completeness limit.  
By comparing the 2-D
distribution of cluster masses and gas surface densities, we find for
clusters older than 25 Myr that $M_{\rm 3rd} \propto \Sigma_{\rm
  HI}^{0.4 \pm 0.2}$, where $M_{\rm 3rd}$ is the median of the 5 most
massive clusters.  There is no correlation with $\Sigma_{\rm gas}$,
$\Sigma_{\rm H2}$, or $\Sigma_{\rm SFR}$. For clusters younger than 10
Myr, $M_{\rm 3rd} \propto \Sigma_{\rm HI}^{0.6 \pm 0.1}$, $M_{\rm 3rd}
\propto \Sigma_{\rm gas}^{0.5 \pm 0.2}$; there is no correlation with
either $\Sigma_{\rm H_2}$ or $\Sigma_{\rm SFR}$.  The results could
hardly be more different than those found for clusters younger than 25
Myr in M~33.  For the flocculent galaxy M~33, there is no correlation
between maximum cluster mass and neutral gas, but we have determined
$M_{\rm 3rd} \propto \Sigma_{\rm gas}^{3.8 \pm 0.3};$ $M_{\rm 3rd}
\propto \Sigma_{\rm H_2}^{1.2 \pm 0.1};$ $M_{\rm 3rd} \propto
\Sigma_{\rm SFR}^{0.9 \pm 0.1}$. For the older sample in M~51, the
lack of tight correlations is probably due to the combination of the
strong azimuthal variations in the surface densities of gas and star
formation rate, and the cluster ages. These two facts mean that
neither the azimuthal average of the surface densities at a given
radius, nor the surface densities at the present-day location of a
stellar cluster represent the true surface densities at the place and
time of cluster formation.  In the case of the younger sample, even if
the clusters have not yet traveled too far from their birthsites, the
poor resolution of the radio data compared to the physical sizes of
the clusters results in measured $\Sigma$s that are likely quite
diluted compared to the actual densities relevant for the formation of
the clusters.

\end{abstract}


\section{Introduction}

This is the second in a series of papers devoted to
investigate whether the maximum young cluster masses observed in a galaxy 
reflect physical processes that determine them or if, instead, 
they constitute an effect of random sampling statistics.

A correlation between gas surface density and embedded maximum cluster mass 
(i.e., $M_{\rm ecl,max} \propto \Sigma_{\rm gas}^{3/2}$) has been proposed as an ansatz 
to explain the existence of the H$\alpha$ cut-off in disk galaxies,
while at the same time, as can be inferred from UV data, there is no actual 
star formation cut-off 
\citep{pfla08,pfla09}. Similarly, a correlation 
$M_{\rm ecl,max} \propto \Sigma_{\rm SFR}^\eta$
has been proposed by \citet{bill02} and \citet{lars02},
arising from the star formation law \citep[$\Sigma_{\rm SFR} 
\propto \Sigma_{\rm gas}^{1.4}$][]{kenn98}, in conjunction with 
the assumption of pressure equilibrium
between the ambient ISM and the cluster-forming cloud cores. In this case,
$2/3 \leq \eta \leq 2$, with the lower value expected for 
equal (volume) density clusters, and the higher one for clusters with equal size \citep{lars02}.

In the first paper of the series \citep{gonz12}, we analyzed 
radially averaged distributions of gas surface densities and 
of cluster masses in the flocculent
galaxy M~33. Here, we study the grand-design galaxy M~51.
Star cluster data are presented in Section 1, and
ISM data in Section 2; both cluster and gas data
are taken from the literature. The results and a comparison
with M~33 are discussed in Section 3. Section 4 contains
our summary and conclusions.

\section{Star cluster data}
\label{scd}

We rely on two catalogs of star clusters brighter
than $V_{F555W}$ = 23 mag, detected by \cite{hwan08},
based on Hubble Space Telescope 
({\sl HST}) Advanced Camera for Surveys
(ACS) observations acquired by the Hubble Heritage Team.
\citeauthor{hwan08} compiled two catalogs, one 
of 2224 ``Class 1" clusters, that appear circular
and well isolated, and one with 1388 ``Class 2" clusters,
which have elongated or irregular shapes and/or close 
neighbors. These original catalogs list, among other properties, 
cluster RA and DEC, half-light radius $R_{\rm eff}$, ($B - V$) 
and ($V - I$) colors, aperture corrected $V$ mag,
and photometric errors, all in the {\sl HST} equivalent of the Johnson system.
From the recovery of artificial clusters,
\citet{hwan08} determined that the completeness of clusters
with $V < 23$ mag is higher than 80\%, regardless of the
clustering properties of the field.

Subsequently, \citet{hwan10} 
calculated masses, ages, and reddening for the clusters,
via the comparison of their broad band\footnote{
The Hubble Heritage ACS data set for M~51 comprises images in the
$F435W$, $F555W$, and $F814W$ filters. \citet{hwan10} complemented
these with archival $F336W$ Wide Field Planetary Camera 2 (WFPC2) data.  
} spectral energy distributions (SEDs)
with theoretical SEDs derived from the \citet{bruz03} stellar population
synthesis models. The models used by \citet{hwan10} have a Salpeter initial mass function (IMF), with
lower and upper mass cutoffs of 0.1 $M_\odot$ and 100 $M_\odot$, respectively; 
ages between 1 Myr and 15 Gyr; and solar metallicity. 
Each one of the 182 model spectra was reddened by $E(B-V)$ values between 0.0 and 0.6 in 40 0.015 mag
steps.\footnote{This range in color excess includes the values obtained for
M~51 clusters by previous studies; most clusters have
$E(B-V) < 0.1$ \citep{lee05}.} The age and reddening of each cluster were determined simultaneously, 
from a $\chi^2$ fit of the reddened model SEDs to the photometric data. 
The original catalogs were thus reduced to 1125 Class 1 and 835 Class 2 objects 
with acceptable fits.
For each cluster, a mass was finally assigned from the mass to light ratio $M/L_V$ 
of the best-fit model and the $V$-band cluster magnitude, taking the measured extinction into
account and adopting a distance to M~51 of 8.4 Mpc \citep{feld97}.\footnote{
At this distance, 1$^{\prime\prime} =$ 40.7 pc; $R_{25}$, the galactocentric distance of
the isophote with surface brightness in the $B$-band $\mu_{B} = 25$ mag $\sq^{\prime\prime -1}$,
is 13.7 kpc or $5\farcm6$.} The \citet{hwan10} catalogs list, for
each cluster, log age and errors, log mass, color excess and errors,
and $\chi^2$ value of the best-fit. These catalogs do not provide an error 
in the mass determination. One possible estimate is given by 
the $M/L_V$ ratios at the minimum and maximum ages (the age 
minus and plus its error, respectively) of each cluster. Except for 24 
(out of 1960) 
clusters (half of them less massive than $5 \times 10^3 M_\odot$),
this error is smaller than
0.15 dex (35\% of the mass) in all cases. 
The dominant source of error is probably the uncertainty in
the distance modulus (0.6 mag, or a factor of $\lesssim$ 2 in mass).
It would affect all the masses equally, though, so it would not
change the exponent of any relationship between 
cluster mass and gas or SFR surface density.

The age distribution of the clusters
shows three peaks, that likely signpost bursts of enhanced 
star formation rate, respectively, at stellar ages of $\approx$ 5, 100, and 200 Myr.
The two older starbursts are better traced by the detected Class 1 objects, 
which have masses ranging from $\approx 3 \times 10^3$ to $\gtrsim 1 \times 10^6 M_\odot$. 
Conversely, detected clusters in the recent burst belong mostly to Class 2, with 
significantly lower masses,
between $\approx 600$ and $4 \times 10^4 M_\odot$.

\section{ISM data}

Since M~51 is a grand-design spiral with large azimuthal variations
of gas surface density, a two-dimensional approach is required, as opposed to
a comparison of individual cluster masses with radial gas profiles,
that was adequate for M~33. 

We use the HI map of M~51 data from \citet{rots90}; 
it was originally acquired with the Very Large Array (VLA) and 
has a spatial resolution of 13$^{\prime\prime}$.
We derive the molecular gas surface density distribution from the 
$^{12}$CO 2--1 mosaic obtained by \citet{schu07} with the 
IRAM-30 m telescope; it has a resolution
of 11$^{\prime\prime}$.
Finally, to calculate the SFR surface density distribution we 
rely on the total power 20 cm radio continuum image published by \citet{flet11}; 
it has a resolution of 15$^{\prime\prime}$, and 
was produced from VLA C- and D-array data.

In order to get the H$_2$, HI, and SFR surface
densities from the integrated intensity maps, we follow the same procedures as
\citet{schu07}. 
At a given position in M~51, the neutral gas surface density $\Sigma_{\rm HI} = m_{\rm H} N$(HI) cos\ $i$, where
$m_{\rm H}$ is the hydrogen atomic mass and $i = 20\degr$ is the
inclination angle of M~51; assuming optically thin 
emission, $N$(HI) = $1.82 \times 10^{18} 
\int T_{\rm mb} {\rm d}v$/(K km s$^{-1}$)cm$^{-2}$. 
Likewise, we obtain the molecular gas surface density as
$\Sigma_{\rm H_2} = 2~m_{\rm H}N$(H$_2$) cos\ $i$. 
\citet{schu07}  assume a
CO 2-1/1-0 intensity
of 0.8, and a constant CO to H$_2$ conversion factor, $X$, 1/4 that of the
Milky Way (MW), with $X_{MW} = 2.3 \times 10^{20}$ cm$^{-2}$
(K km s$^{-1}$)$^{-1}$, such that the molecular hydrogen column density is   
$N$(H$_2$) = 0.25 $X_{\rm MW} (1/0.8) \int T_{\rm mb}(\rm CO (2-1))$d$v$,
where $T_{\rm mb}$ is the main beam brightness temperature. 
The total gas surface density is $\Sigma_{\rm gas} = 1.36(\Sigma_{\rm H_2} + \Sigma_{\rm HI})$,
so as to include helium. 
Again, following \citet{schu07}, we derive $\Sigma_{\rm SFR}$ from the 20 cm 
radio continuum data.\footnote{$\Sigma_{\rm SFR} = 1.53 \times 10^5  
\frac{S_{\rm 20 cm} {\rm cos} (i)}{[\rm Jy\ beam^{-1} ]} M_\odot {\rm pc^{-2} Gyr^{-1}}$.}
The 20 cm non-thermal radio continuum is due to synchrotron radiation by
supernovae-accelerated cosmic rays; it 
hence traces recent star formation,
and has a strong correlation with the far-infrared
\citep[FIR; e.g.,][]{helo85}.\footnote{ 
$\Sigma_{\rm SFR}$ can be readily derived from FIR emission,
given that a significant amount of it is due to reprocessing by dust of light from young stars; 
FIR emission also has the advantage that it does not 
need to be corrected for extinction. In fact, \citet{rosa02} find that
SFRs obtained at optical and UV
wavelengths are often overcorrected for extinction, since stellar Balmer absorption
contaminates emission lines; optical/UV and FIR/mm SFR estimators agree well
once this effect is taken into account.} 
At the position of each star cluster, gas surface densities
are obtained from the HI, H$_2$, and 20 cm mosaics averaging, respectively, the 
emission over $20 \times 20 \sq^{\prime\prime}$ ($\approx 820 \times 820$ pc$^2$), 
$18 \times 18 \sq^{\prime\prime}$ 
($\approx 730 \times 730$ pc$^2$), 
and $18 \times 18 \sq^{\prime\prime}$.\footnote{
The limit is imposed by the pixel size of the CO image, i.e.,
6$^{\prime\prime}$ on the side; the smallest area over which we can
average the flux is a subarray of 3$\times$3 pixels 
(sides of subarrays must have an odd number of pixels), centered on the position of each
cluster. We match the elements over which we average the HI and 20 cm fluxes to this 
$18^{\prime\prime} \times 18^{\prime\prime}$ footprint
as closely as is allowed by the pixel sizes of the images, 
respectively, $4^{\prime\prime} \times 4^{\prime\prime}$ (arrays with 
5 pixels on the side) 
and $2^{\prime\prime} \times 2^{\prime\prime}$ (arrays of 9 $\times$ 9 square 
pixels).}
Ninety-five of the clusters in the sample have coordinates that fall outside of 
the borders of the radio images, so that only 1865 objects 
can be compared with the gas and SFR densities at their location in the 
M~51 disk. 

\section{Results}

\citet{hwan08} find that the completeness 
distribution is quite similar for both Class 1 and Class 2 clusters. 
Since cluster brightness in the optical diminishes with age, the completeness
limit of the cluster mass function depends on cluster age. 
Based on their photometric completeness limits, \citeauthor{hwan08} determine that 
conservative limits over which their mass functions are not 
affected by incompleteness are $\approx 5 \times 10^3 M_\odot$ for clusters younger than
10$^7$, and at least 1$\times 10^5 M_\odot$ for clusters between 
100 and 250 Myr. 
  
\subsection{The bursts 100 and 250 Myr ago}

We will first 
work with the 167 clusters in the catalogs with at least $10^5 M_\odot$, 
ages between 40 and 400 Myr,\footnote{The young tale of the age distribution of the
clusters in the $10^8$ yr old burst reaches 25 Myr, but no cluster 
in this group that is younger than 40 Myr makes the $10^5 M_\odot$ cut.} 
and detected CO, HI, and 20 cm emission
at their present locations. 
\citet{chan10} and \citet{chan11} find that the cluster mass function is approximately
independent of age (and the age function independent of mass) for clusters younger 
than $4\times 10^8$ yr in M~83
and M~51\citep{chan11}, respectively. Similar results are found for the
Magellanic Clouds \citep{parm08} and the Antena galaxies \citep{fall09}.
The implication is that the mass-dependent disruption timescale is $>$ 2 Gyr
\citep{chan10}.

We take care here to compare cluster mass with gas and SFR surface 
density always in bins with the same number of clusters, in order to 
test whether any change in 
maximum cluster mass is a statistical,
size-of-sample effect.\footnote{
If clusters
are drawn randomly from the same mass distribution function that declines
with mass, the probability of picking a massive cluster increases with the size of sample.}
We search for correlations between 
cluster mass, and gas and SFR surface densities of the form:
\begin{equation}
{\rm log}_{10}\ M_{i} = \beta^\prime_x {\rm log}_{10}\ \Sigma_x + \alpha^\prime_x, 
\end{equation}

\noindent
where $M_i$ is the mass of the $i$-th most massive star cluster, and $x$ stands for
HI, H2, total gas or SFR.

\begin{figure*}
\includegraphics[angle=-90.,width=0.85\hsize,clip=]{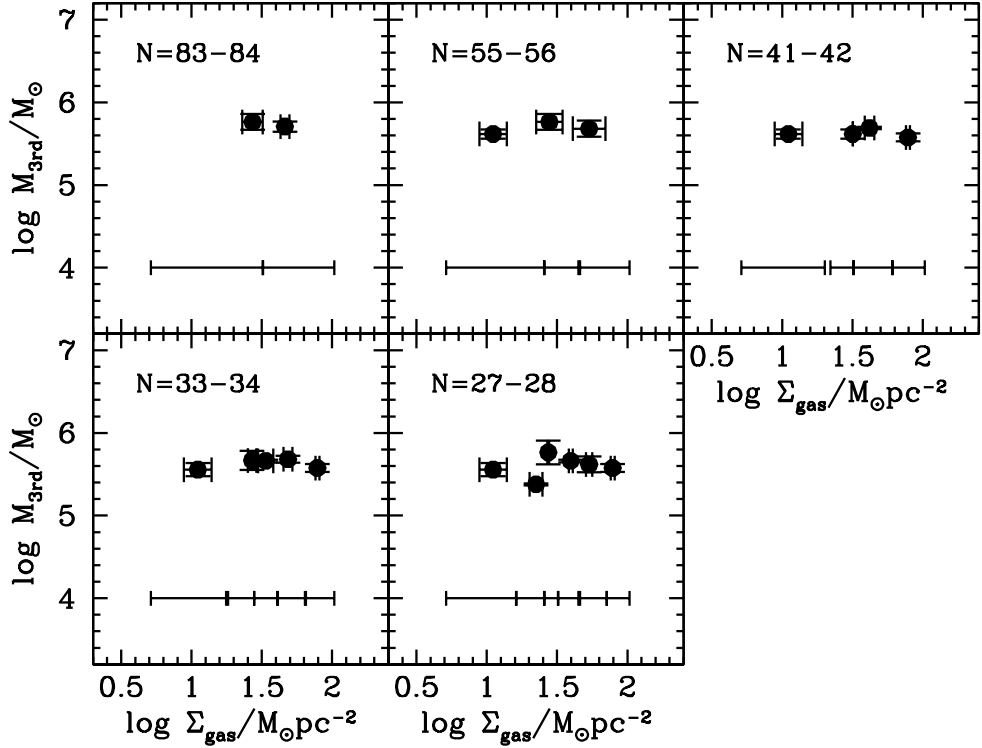}
\caption{M~51, 100 and 250 Myr old bursts, log$_{10}$ cluster mass versus log$_{10}$ total gas surface density, in bins with 
equal number of clusters. 
{\it Filled circles:} $M_{\rm 3rd}$, median of 5 most massive clusters
(i.e., effectively, the third most massive cluster) in the bins. 
Number of bins
increases from 2 (upper left panel) to 6 (lower middle panel); number of 
clusters in each bin are indicated in the panels. 
The bar at the bottom
of each panel shows surface density ranges of the bins. 
}
\label{m51_stot}
\end{figure*}

\begin{figure*}
\includegraphics[angle=-90.,width=0.85\hsize,clip=]{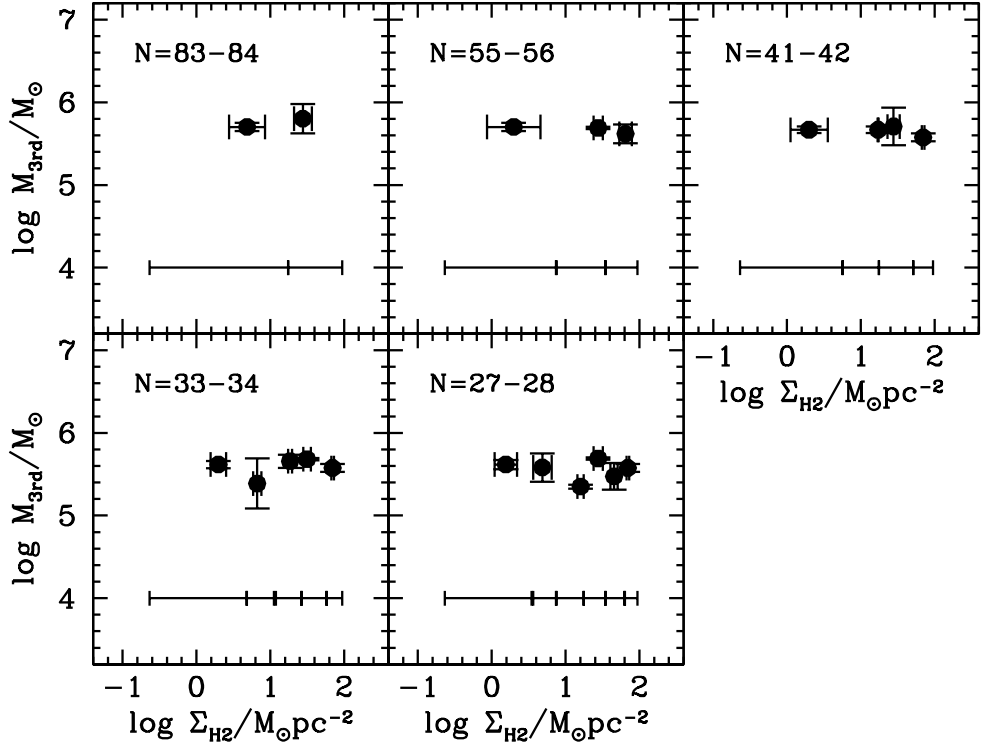}
\caption{M~51, 100 and 250 Myr old bursts, log$_{10}$ cluster mass versus log$_{10}$ molecular gas surface density. 
Symbols as in figure~\ref{m51_stot}.}
\label{m51_sH2}
\end{figure*}

\begin{figure*}
\includegraphics[angle=-90.,width=0.85\hsize,clip=]{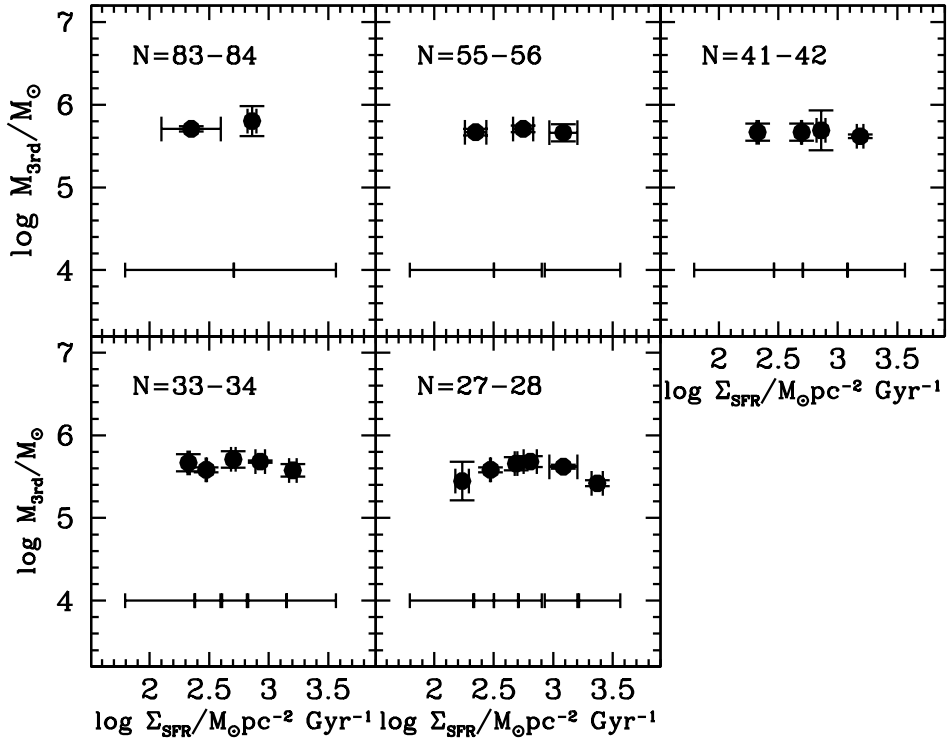}
\caption{M~51, 100 and 250 Myr old bursts, log$_{10}$ cluster mass versus log$_{10}$ star formation rate surface density. 
Symbols as in figure~\ref{m51_stot}.}
\label{m51_sSFR}
\end{figure*}

\begin{figure*}
\includegraphics[angle=-90.,width=0.85\hsize,clip=]{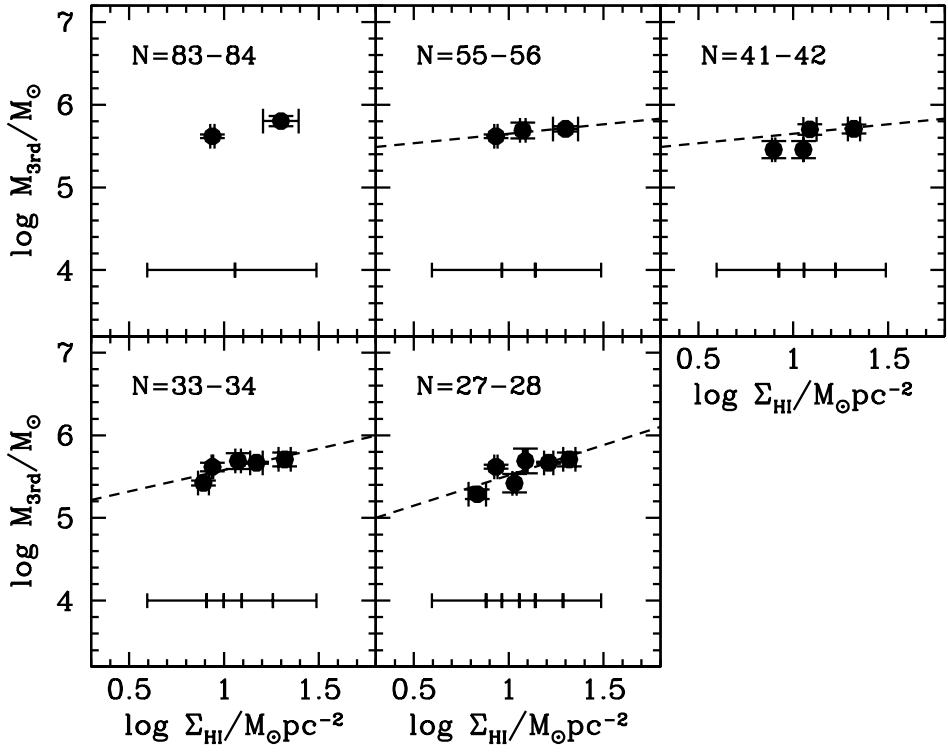}
\caption{M~51, 100 and 250 Myr old bursts, log$_{10}$ cluster mass versus log$_{10}$ neutral gas surface density. 
{\it Dashed line:} linear fit. 
Other symbols as in figure~\ref{m51_stot}.}
\label{m51_sHI}
\end{figure*}

Figures~\ref{m51_stot},~\ref{m51_sH2},~\ref{m51_sSFR}, and~\ref{m51_sHI} 
show log of star cluster mass versus, respectively,
log of total gas, molecular gas, star formation rate, and neutral gas surface
densities.  
The comparison is performed in bins with equal number of clusters. 
Bins increase from 2 (upper left panel) to 6 (lower middle panel), 
and the number of clusters in each bin is indicated. 
Instead of using the maximum cluster mass in each bin for the
comparison, the median of the five most massive clusters 
(or the third most massive cluster, $M_{\rm 3rd}$) in each bin
is displayed as a filled circle; the error bars represent the
interquartile range, for both mass and gas/SFR surface
density. Since the uncertainty in the measurement
of any individual mass is typically a factor of 2-3, the 
median of the 5 most massive clusters in each bin should in general 
be a more statistically robust
indicator of any existing trend
(in every bin there are always more than
5 clusters more massive than the completeness limit).
There is no correlation of $M_{\rm 3rd}$ with the surface densities of 
total gas, $\Sigma_{\rm gas}$, molecular gas, $\Sigma_{\rm H2}$, or
star formation rate, $\Sigma_{\rm SFR}$.

Conversely, there is a hint of a correlation with $\Sigma_{\rm HI}$.
The slopes ($\beta^{\prime}_{\rm HI}$) and intercepts ($\alpha^{\prime}_{\rm HI}$) of the
fits in the bins with 3 or more points, as well as
their uncertainties, are listed in Table 1. 
The fits' average can be expressed as log$_{10}\ M_{\rm 3rd} = 
(0.4 \pm 0.2)\ {\rm log}_{10}\ \Sigma_{\rm HI} + (5.2 \pm 0.3)$. 

\begin{deluxetable}{lrrrrrr}
\tabletypesize{\small}
\tablecolumns{4}
\tablewidth{15cm}
\tablecaption{M~51, 100 and 250 Myr old bursts. Fits to log$_{10}\ (M_{\rm 3rd}/M_\odot)$ vs.\ log$_{10}\ \Sigma_{\rm HI}$/($M_\odot$ pc$^{-2}$)}
\tablehead{
\colhead{$N_b$}&
\colhead{$N_{\rm cl}$}&
\colhead{$\beta^{\prime}_{\rm HI}$}&
\colhead{$\alpha^{\prime}_{\rm HI}$}\\
}
\startdata
 3 & 55-56& $(0.2 \pm 0.1) $ & $(5.4 \pm 0.1)$ & \\
 4 & 41-42& $(0.2 \pm 0.1)$ & $(5.4 \pm 0.1)$  \\
 5 & 33-34& $(0.5 \pm 0.2)$ & $(5.1 \pm 0.3)$  \\
 6 & 27-28& $(0.7 \pm 0.3)$ & $(4.8 \pm 0.3)$  \\
\enddata
\tablecomments{Col.\ (1): number of bins. Col.\ (2):
number of clusters in each bin. Col.\ (3): best-fit slope (eq.~1).
Col.\ (4): best-fit intercept (eq.~1). 
}
\end{deluxetable}

\begin{figure*}
\includegraphics[angle=0.,width=1.00\hsize,clip=]{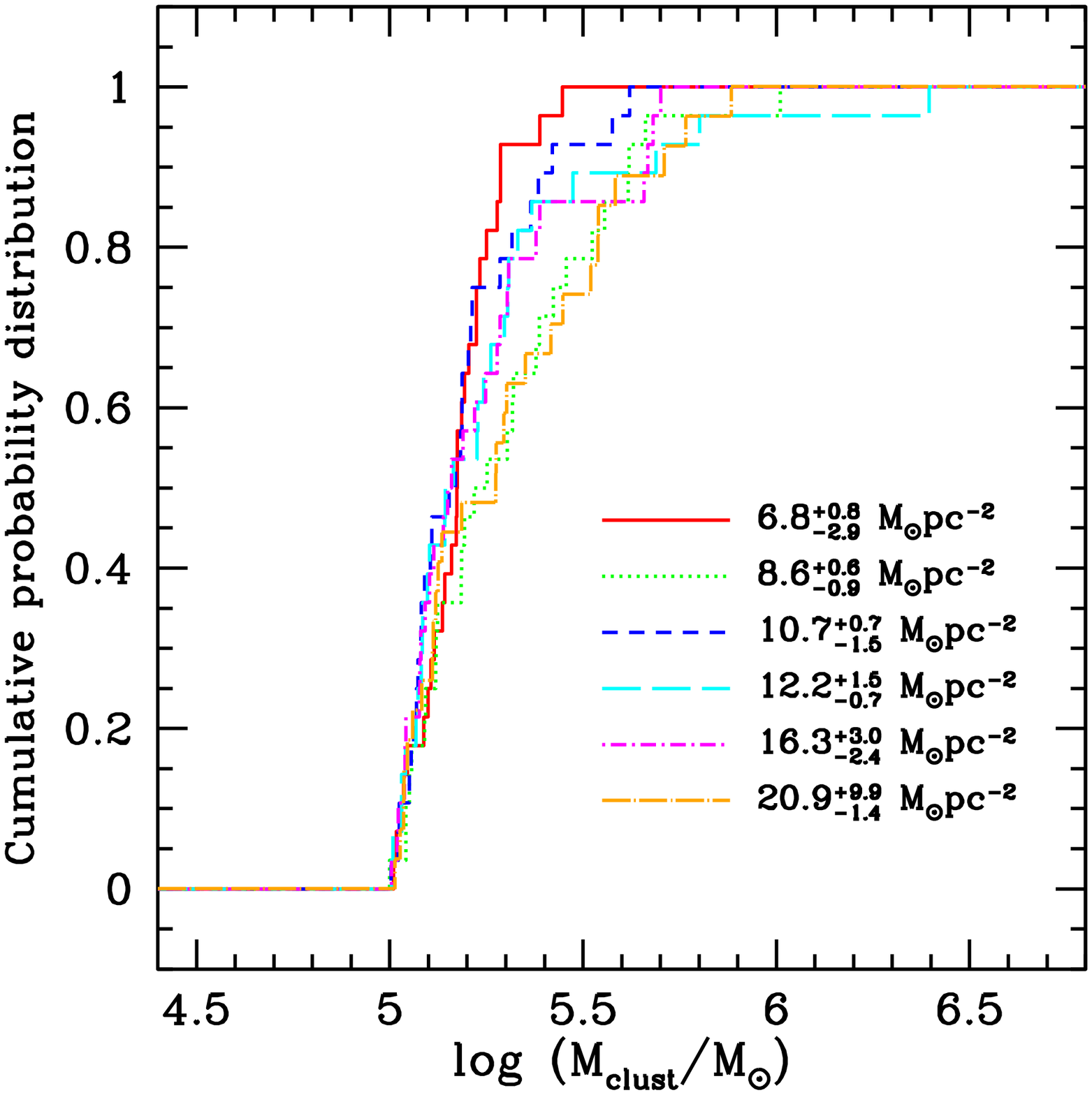}
\caption{K-S test, 100 and 250 Myr old bursts. Cumulative probability distributions
of mass for clusters with $M_{\rm clust} \geq 10^5\ M_\odot$ in the six
indicated bins of HI surface density. {\it Red solid line:}
bin 1, $\Sigma_{\rm HI} = 6.8^{+0.8}_{-2.9}\ M_\odot$ pc$^{-2}$;
{\it green dotted line:} bin 2, $\Sigma_{\rm HI} = 8.6^{+0.6}_{-0.9}\ M_\odot$ pc$^{-2}$;
{\it blue short dashed line:} bin 3,  $\Sigma_{\rm HI} = 10.7^{+0.7}_{-1.5}\ M_\odot$ pc$^{-2}$;
{\it cyan long dashed line:} bin 4,  $\Sigma_{\rm HI} = 12.2^{+1.5}_{-0.7}\ M_\odot$ pc$^{-2}$;
{\it magenta short dashed-dotted line:} bin 5, $\Sigma_{\rm HI}= 16.3^{+3.0}_{-2.4}\  M_\odot$ pc$^{-2}$;
{\it orange long dashed-dotted line:} bin 6, $\Sigma_{\rm HI}= 20.9^{+9.9}_{-1.4}\ M_\odot$ pc$^{-2}$.
}
\label{m51_ks}
\end{figure*}

In order to further explore whether the intrinsic mass distribution of
clusters may be changing as a function of neutral gas surface density, we perform a Kolmogorov-Smirnov
(K-S) test on the 6 different subsamples presented in the bottom
middle panel of Figure~\ref{m51_sHI}. Figure~\ref{m51_ks} shows the cumulative probability
distributions and $\Sigma_{\rm HI}$ of each subsample, and the $D_{j,k}$ and $P_{j,k}$ values
for every bin pair $j,k$ are given in Table 2 (higher bin number indicates larger
surface density). 
Assuming that sample pairs with $P_{j,k} < 0.05$ are
taken from different distribution functions with
high significance, bin 1 ({\it red solid line}), with the smallest surface density 
$\Sigma_{\rm HI} = 6.8^{+0.8}_{-2.9} M_\odot$ pc$^{-2}$, 
seems to be different from bin 6 ({\it orange long dashed-dotted line}), with the largest surface density
$\Sigma_{\rm HI} = 20.9^{+9.9}_{-1.4} M_\odot$ pc$^{-2}$ ($P_{1,6} = 0.03$).  
The mass distributions in bins 3, 4, and 5 appear to lie somewhere in
between. No hard conclusion can be drawn, however, since bin 2 ({\it green dotted line}), with 
the second lowest HI surface density, is 
in fact indistinguishable from bin 6 ($P_{2,6} > 0.98$ for the pair).

\begin{table*}
\center{\sc Table 2\\ K-S test $D$ and $P$ values, 100 and 250 Myr old bursts}\\
\center{}
  \hspace*{1.25cm}
 \begin{minipage}{140mm}
\begin{small}
\begin{tabular}{@{}c|cc|cc|cc|cc|cc@{}}
\hline
\hline
\vspace*{-0.021cm}& \vspace*{-0.021cm}& \vspace*{-0.021cm}&\vspace*{-0.021cm}& \vspace*{-0.021cm} &\vspace*{-0.021cm} &\vspace*{-0.021cm} &\vspace*{-0.021cm} &\vspace*{-0.021cm} &\vspace*{-0.021cm} &\vspace*{-0.021cm}  \\
 & $D$ & $P$ & $D$ & $P$ & $D$ & $P$ & $D$ & $P$ & $D$ & $P$  \\
\vspace*{-0.026cm}& \vspace*{-0.026cm}& \vspace*{-0.026cm}&\vspace*{-0.026cm}& \vspace*{-0.026cm} &\vspace*{-0.026cm} &\vspace*{-0.026cm} &\vspace*{-0.026cm} &\vspace*{-0.026cm} &\vspace*{-0.026cm} &\vspace*{-0.026cm} \\ \hline
\multicolumn{1}{c}{Bin}&\multicolumn{2}{|c|}{2}&\multicolumn{2}{|c|}{3}&\multicolumn{2}{|c|}{4}&\multicolumn{2}{|c|}{5}&\multicolumn{2}{|c}{6} \\ \hline
1 & 1.53 & 0.0187 & 0.69 & 0.7198 & 0.97 & 0.3003 & 0.83 & 0.4903 & 1.44  & 0.0318   \\ \hline
2 &      &        & 1.11 & 0.1687 & 0.83 & 0.4903 & 0.83 & 0.4903 & 0.46  & 0.9823   \\ \hline
3 &      &        &      &        & 0.83 & 0.4903 & 0.69 & 0.7198 & 1.04  & 0.2337   \\ \hline
4 &      &        &      &        &      &        & 0.28 & 1.0000 & 0.76  & 0.6100   \\ \hline
5 &      &        &      &        &      &        &      &        & 0.73  & 0.6530  \\
\hline
\tablecomments{\small $D$ and $P$ values for bin pairs. The cell in the intersection
of a row $j$ and a column $k$ contains the $D_{j,k}$ and $P_{j,k}$
parameters, respectively, of the comparison between the two bins indicated in the
corresponding row and column. If $P_{j,k} < 0.05$, the null hypothesis that the
clusters in the two bins are taken from the same mass distribution function is rejected.}
\end{tabular}
\end{small}
\end{minipage}
\label{kstab}
\end{table*}

\subsection{The recent burst 5 Myr ago}

\begin{figure*}
\includegraphics[angle=-90.,width=0.85\hsize,clip=]{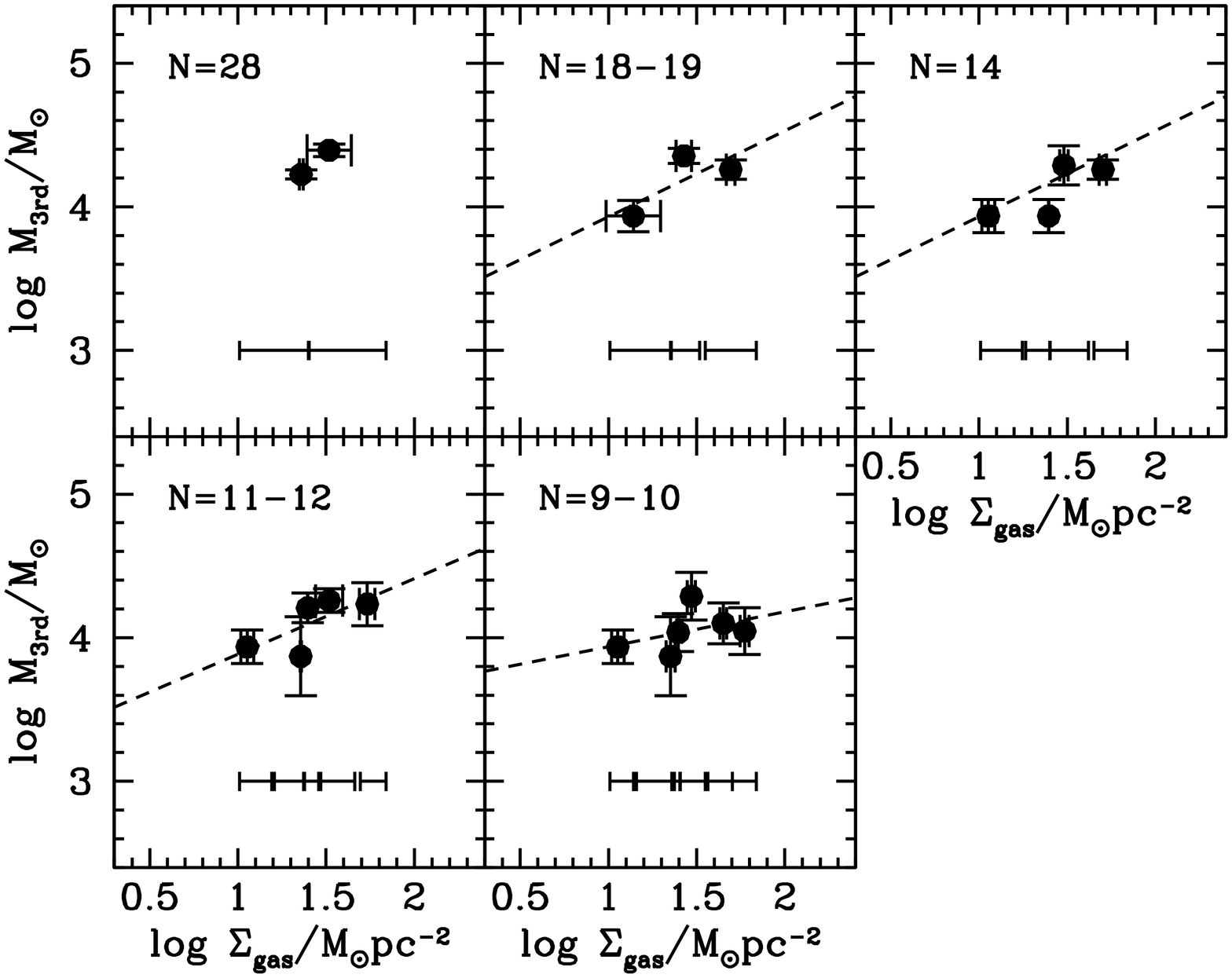}
\caption{M~51, burst 5 Myr ago, log$_{10}$ cluster mass versus log$_{10}$ total gas surface density, in bins with equal number of clusters.
Symbols as in figure~\ref{m51_stot}.
}
\label{m51_young_stot}
\end{figure*}

\begin{figure*}
\includegraphics[angle=-90.,width=0.85\hsize,clip=]{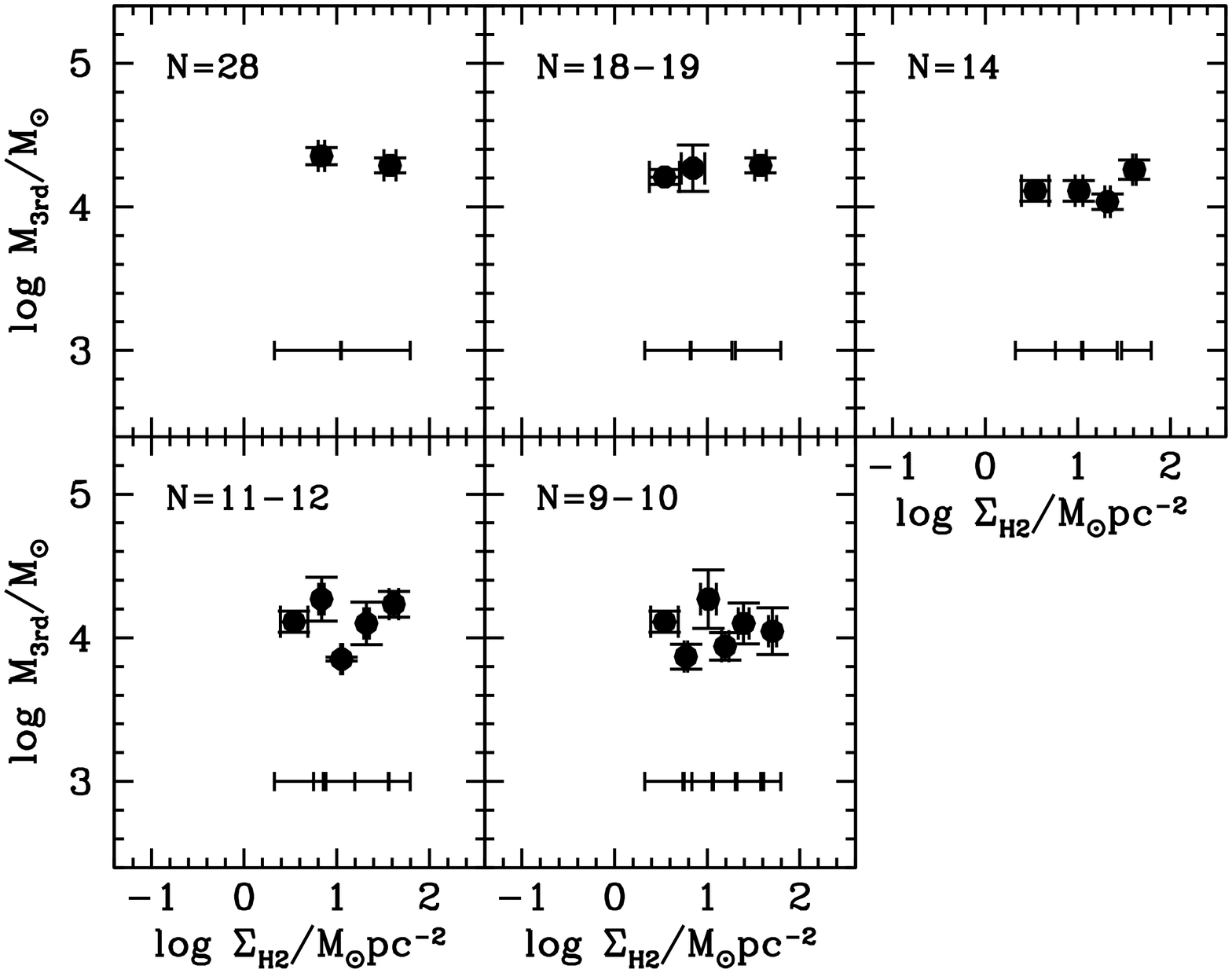}
\caption{M~51, burst 5 Myr ago, log$_{10}$ cluster mass versus log$_{10}$ molecular gas surface density.
Symbols as in figure~\ref{m51_stot}.}
\label{m51_young_sH2}
\end{figure*}

\begin{figure*}
\includegraphics[angle=-90.,width=0.85\hsize,clip=]{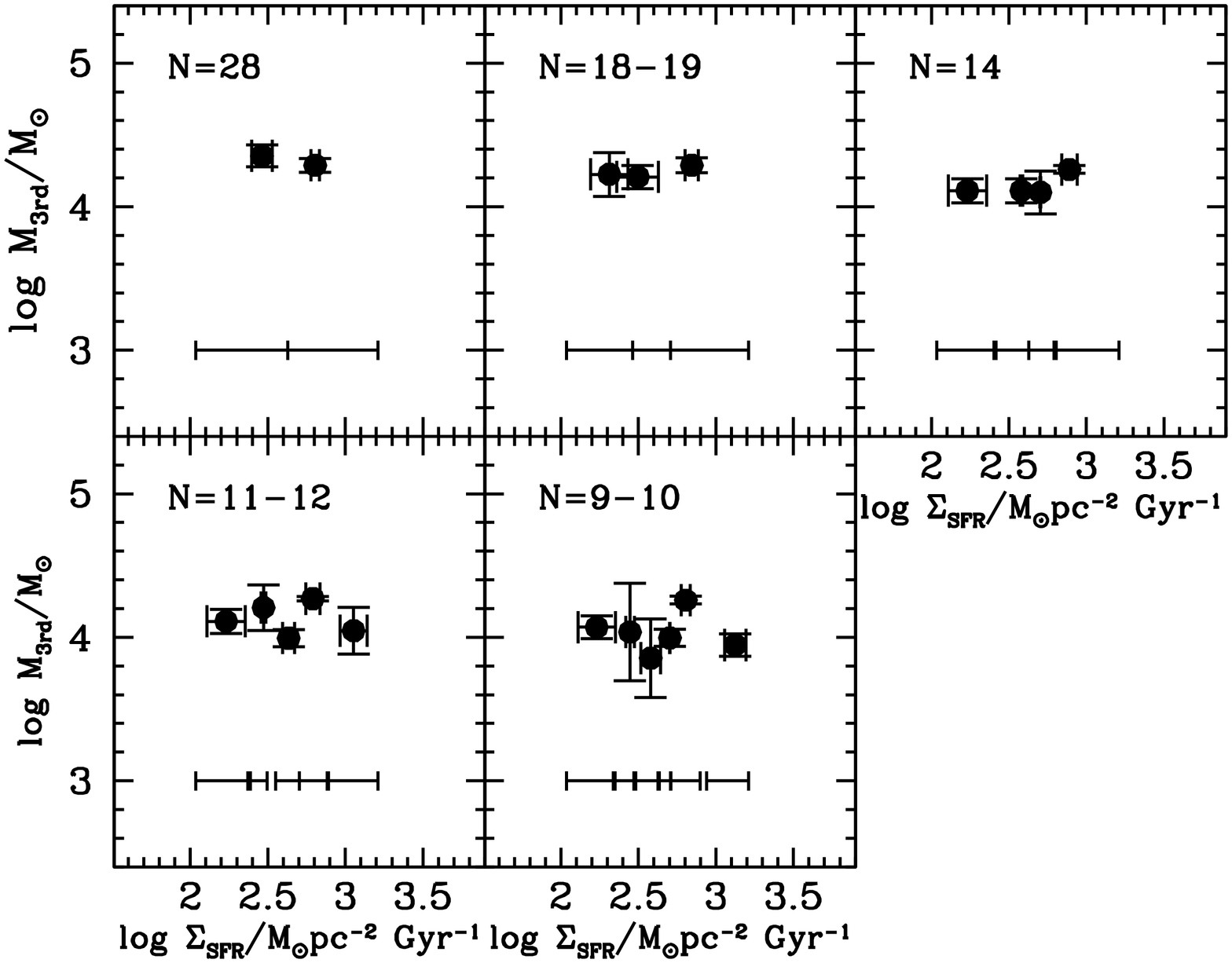}
\caption{M~51, burst 5 Myr ago, log$_{10}$ cluster mass versus log$_{10}$ star formation rate surface density.
Symbols as in figure~\ref{m51_stot}.}
\label{m51_young_sSFR}
\end{figure*}

\begin{figure*}
\includegraphics[angle=-90.,width=0.85\hsize,clip=]{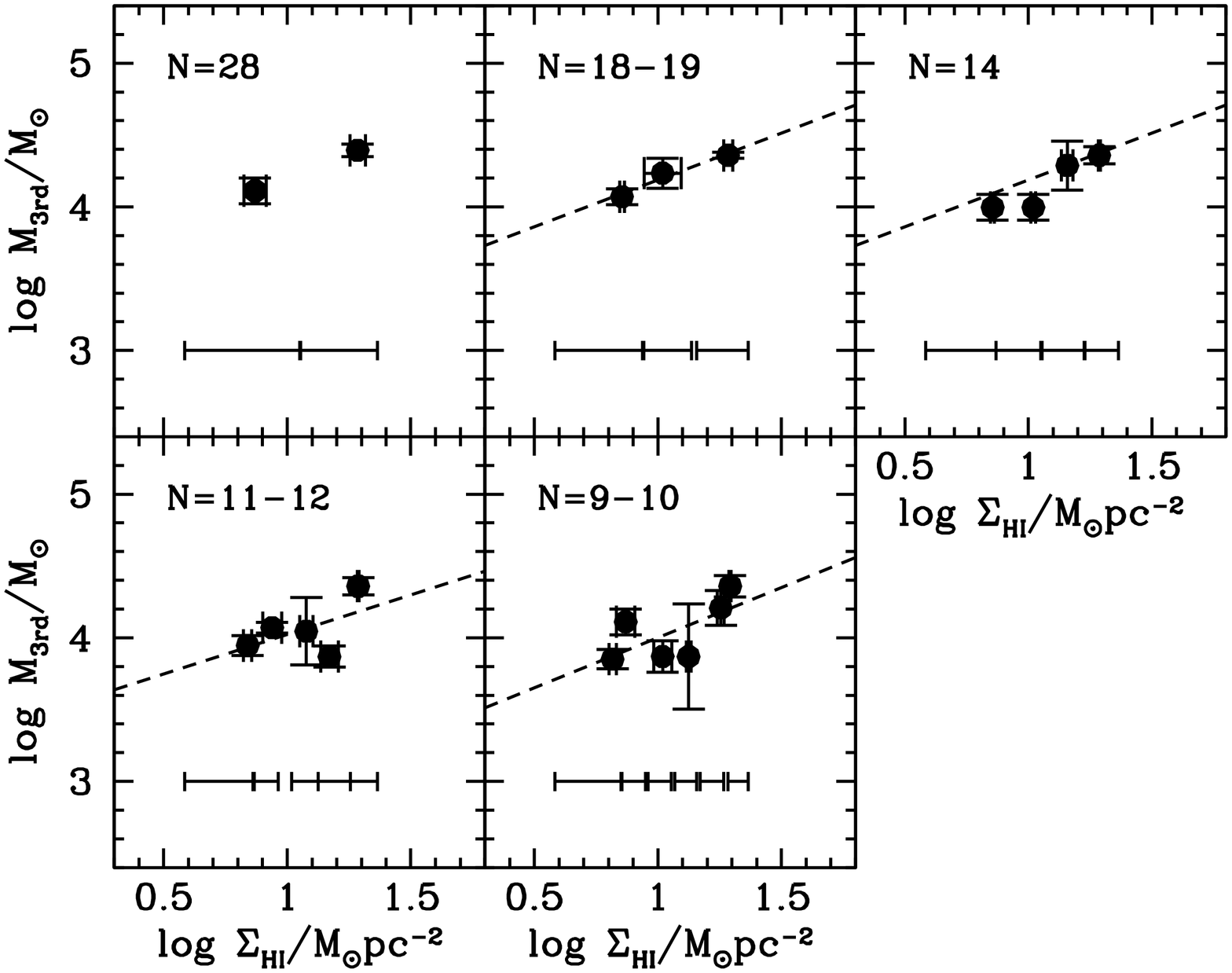}
\caption{M~51, burst 5 Myr ago, log$_{10}$ cluster mass versus log$_{10}$ neutral gas surface density.
{\it Dashed line:} linear fit.
Other symbols as in figure~\ref{m51_stot}.}
\label{m51_young_sHI}
\end{figure*}

We now analyze the 56 clusters younger than 10$^7$ yr with mass at least
5$\times 10^3 M_\odot$ and higher, and detected CO, HI, and 20 cm emission
at their positions. 
Figures~\ref{m51_young_stot},~\ref{m51_young_sH2},~\ref{m51_young_sSFR}, and~\ref{m51_young_sHI} 
show log of star cluster mass versus, respectively,
log of $\Sigma_{\rm gas}$, $\Sigma_{\rm H2}$, $\Sigma_{\rm SFR}$, and
$\Sigma_{\rm HI}$.
Bins with indicated, equal, number of clusters again increase from 2 to 6; points
are the median of the 5 most massive clusters in each bin. 

\begin{figure*}
\includegraphics[angle=-90.,width=0.85\hsize,clip=]{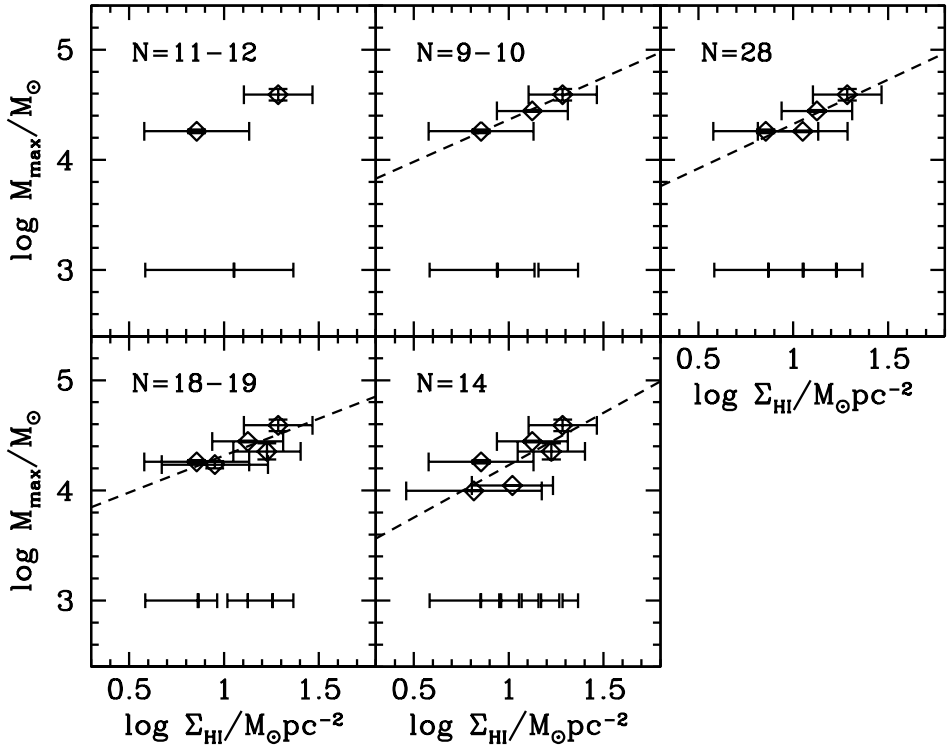}
\caption{M~51, burst 5 Myr ago, log$_{10}$ cluster mass 
versus log$_{10}$ neutral gas surface density.
{\it Empty diamonds:} maximum cluster mass; {\it dashed line:} linear fit.
Other symbols as in figure~\ref{m51_stot}.}
\label{m51_young_max_sHI}
\end{figure*}

As with clusters in the two older bursts, there is no correlation
between $M_{\rm 3rd}$, and  
$\Sigma_{\rm H2}$ or $\Sigma_{\rm SFR}$.
The correlation with $\Sigma_{\rm HI}$ is tighter and has a slightly steeper slope, 
though, with average
fit for all the bins with 3 or more points 
log$_{10}\ M_{\rm 3rd} = (0.6 \pm 0.1)\ {\rm log}_{10}\ \Sigma_{\rm HI} 
+  (3.5 \pm 0.1)$. Moreover, there is also a hint of a correlation of $M_{\rm 3rd}$ with
the total mass surface density $\Sigma_{\rm gas}$, with average fit
log$_{10}\ M_{\rm 3rd} = (0.5 \pm 0.2)\ {\rm log}_{10}\ \Sigma_{\rm gas} 
+  (3.4 \pm 0.2)$. 
Fully aware that the statistical significance might be questionable,
we show nontheless log$_{10}$ of {\em maximum} cluster mass in the bin, $M_{\rm max}$, versus 
$\Sigma_{\rm HI}$ for clusters younger than 10$^7$ yr in Figure~\ref{m51_young_max_sHI}.
The error in log mass for each cluster was determined as described previously in 
Section~\ref{scd}; the error bar in $\Sigma_{\rm HI}$ shows the
uncertainty in the measurement at the cluster's position. 
The average of the fits in the bins with more than 3 
points is as tight as for $M_{\rm 3rd}$, and consistent with it:
log$_{10}\ M_{\rm max} = (0.7 \pm 0.1)\ {\rm log}_{10}\ \Sigma_{\rm gas}  
+  (3.6 \pm 0.1)$. 
The parameters of the fits for the younger burst 
can be found in Table 3. 

\begin{deluxetable}{lrrrrrr}
\tablenum{3}
\tabletypesize{\small}
\tablecolumns{4}
\tablewidth{15cm}
\tablecaption{M~51, burst 5 Myr ago. Fits to log$_{10}\ (M/M_\odot)$ vs.\ log$_{10}\ \Sigma$, $\beta^{\prime}_x$ and $\alpha^{\prime}_x$ }
\tablehead{
\colhead{$N_b$}&
\colhead{$N_{\rm cl}$}&
\colhead{$\beta^{\prime}_{\rm HI}$}&
\colhead{$\alpha^{\prime}_{\rm HI}$}\\
}
\startdata
\multicolumn{4}{c}{log$_{10}\ (M_{\rm 3rd}/M_\odot)$ vs $\Sigma_{\rm HI}/M_\odot$ pc$^{-2}$} \\ \hline
 3 & 18-19& $(0.7 \pm 0.1) $ & $(3.5 \pm 0.1)$ & \\
 4 & 14& $(0.7 \pm 0.1)$ & $(3.5 \pm 0.1)$  \\
 5 & 11-12& $(0.6 \pm 0.5)$ & $(3.5 \pm 0.6)$  \\
 6 & 9-10& $(0.7 \pm 0.4)$ & $(3.3 \pm 0.4)$  \\ \hline
\multicolumn{4}{c}{log$_{10}\ (M_{\rm 3rd}/M_\odot)$ vs $\Sigma_{\rm gas}/M_\odot$ pc$^{-2}$} \\ \hline
 3 & 18-19 & $(0.6 \pm 0.5)$ & $(3.3 \pm 0.8)$  \\
 4 & 14 & $(0.6 \pm 0.5)$ & $(3.3 \pm 0.8)$ \\
 5 & 11-12 & $(0.5 \pm 0.3)$ & $(3.4 \pm 0.4)$  \\ 
 6 & 9-10  & $(0.2 \pm 0.3)$ & $(3.7 \pm 0.4)$   \\ \hline
\multicolumn{4}{c}{log$_{10}\ (M_{\rm max}/M_\odot)$ vs $\Sigma_{\rm HI}/M_\odot$ pc$^{-2}$} \\ \hline
 3 & 18-19& $(0.8 \pm 0.1)$ & $(3.6  \pm 0.1)$   \\
 4 & 14& $(0.8 \pm 0.3)$ & $(3.5 \pm 0.3)$  \\
 5 & 11-12& $(0.7 \pm 0.3)$ & $(3.6 \pm 0.3)$  \\ 
 6 & 9-10  & $(1.0 \pm 0.4)$ & $(3.3 \pm 0.4)$ \\ \hline
\enddata
\tablecomments{Col.\ (1): number of bins. Col.\ (2):
number of clusters in each bin. Col.\ (3): best-fit slope (eq.~1).
Col.\ (4): best-fit intercept (eq.~1). 
}
\end{deluxetable}

\begin{figure*}
\includegraphics[angle=0.,width=1.00\hsize,clip=]{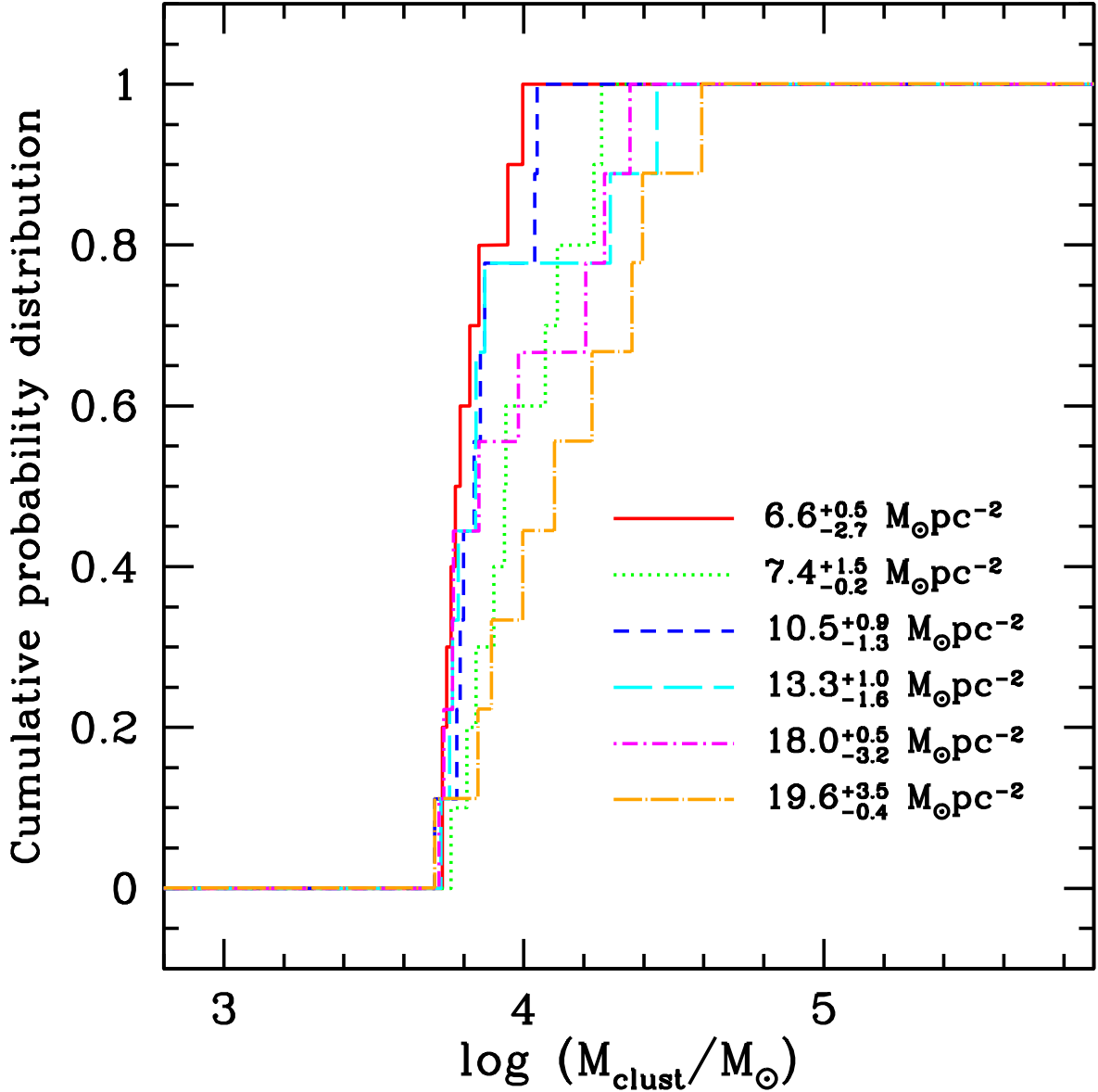}
\caption{K-S test, burst 5 Myr ago. Cumulative probability distributions
of mass for clusters in recent burst with $M_{\rm clust} \geq 5 \times 10^3\ M_\odot$ in the six
indicated bins of HI surface density. {\it Red solid line:}
bin 1, $\Sigma_{\rm HI} = 6.6^{+0.5}_{-2.7}\ M_\odot$ pc$^{-2}$;
{\it green dotted line:} bin 2, $\Sigma_{\rm HI} = 7.4^{+1.5}_{-0.2}\ M_\odot$ pc$^{-2}$;
{\it blue short dashed line:} bin 3,  $\Sigma_{\rm HI} = 10.5^{+0.9}_{-1.3}\ M_\odot$ pc$^{-2}$;
{\it cyan long dashed line:} bin 4,  $\Sigma_{\rm HI} = 13.3^{+1.0}_{-1.6}\ M_\odot$ pc$^{-2}$;
{\it magenta short dashed-dotted line:} bin 5, $\Sigma_{\rm HI}= 18.0^{+0.5}_{-3.2}\  M_\odot$ pc$^{-2}$;
{\it orange long dashed-dotted line:} bin 6, $\Sigma_{\rm HI}= 19.6^{+3.5}_{-0.4}\ M_\odot$ pc$^{-2}$.
}
\label{m51_young_ks}
\end{figure*}

\begin{table*}
\center{\sc Table 4\\ K-S test $D$ and $P$ values, burst 5 Myr ago}\\
\center{}
  \hspace*{1.25cm}
 \begin{minipage}{140mm}
\begin{small}
\begin{tabular}{@{}c|cc|cc|cc|cc|cc@{}}
\hline
\hline
\vspace*{-0.021cm}& \vspace*{-0.021cm}& \vspace*{-0.021cm}&\vspace*{-0.021cm}& \vspace*{-0.021cm} &\vspace*{-0.021cm} &\vspace*{-0.021cm} &\vspace*{-0.021cm} &\vspace*{-0.021cm} &\vspace*{-0.021cm} &\vspace*{-0.021cm}  \\
 & $D$ & $P$ & $D$ & $P$ & $D$ & $P$ & $D$ & $P$ & $D$ & $P$  \\
\vspace*{-0.026cm}& \vspace*{-0.026cm}& \vspace*{-0.026cm}&\vspace*{-0.026cm}& \vspace*{-0.026cm} &\vspace*{-0.026cm} &\vspace*{-0.026cm} &\vspace*{-0.026cm} &\vspace*{-0.026cm} &\vspace*{-0.026cm} &\vspace*{-0.026cm} \\ \hline
\multicolumn{1}{c}{Bin}&\multicolumn{2}{|c|}{2}&\multicolumn{2}{|c|}{3}&\multicolumn{2}{|c|}{4}&\multicolumn{2}{|c|}{5}&\multicolumn{2}{|c}{6} \\ \hline
1 & 1.20 & 0.1108 & 0.91 & 0.3754 & 0.60 & 0.8646 & 0.83 & 0.4892 & 1.38  & 0.0438   \\ \hline
2 &      &        & 1.12 & 0.1617 & 1.12 & 0.1617 & 0.81 & 0.5305 & 0.78  & 0.5732   \\ \hline
3 &      &        &      &        & 0.51 & 0.9575 & 0.76 & 0.6030 & 1.27  & 0.0778   \\ \hline
4 &      &        &      &        &      &        & 0.51 & 0.9575 & 1.27  & 0.0778   \\ \hline
5 &      &        &      &        &      &        &      &        & 0.76  & 0.6030  \\
\hline
\tablecomments{\small $D$ and $P$ values for bin pairs. The cell in the intersection
of a row $j$ and a column $k$ contains the $D_{j,k}$ and $P_{j,k}$
parameters, respectively, of the comparison between the two bins indicated in the
corresponding row and column. If $P_{j,k} < 0.05$, the null hypothesis that the
clusters in the two bins are taken from the same mass distribution function is rejected.}
\end{tabular}
\end{small}
\end{minipage}
\label{kstabyoung}
\end{table*}

We perform once more a 
K-S test on the 6 different subsamples presented in the bottom
middle panel of Figure~\ref{m51_young_sHI}. The test is illustrated
with Figure~\ref{m51_young_ks}, and its numerical results are
given in Table 4.  
Bin 1 ({\it red solid line}), with the smallest surface density
$\Sigma_{\rm HI} = 6.6^{+0.5}_{-2.7} M_\odot$ pc$^{-2}$,
seems to be different, with high statistical
significance, from bin 6 ({\it orange long dashed-dotted line}), with the largest surface density
$\Sigma_{\rm HI} = 19.6^{+3.5}_{-0.4} M_\odot$ pc$^{-2}$ ($P_{1,6} = 0.04$).
The mass distibutions in all the other bins lie in between, although
we notice that the distributions in bins 4 ({\it cyan long dashed line}) and 5 
({\it magenta dashed-dotted line}) are actually closer 
to bin 1 than to bin 6 ($P_{1,4} = 0.86$; $P_{4,6} = 0.08$; $P_{1,5}=0.49$; $P_{5,6} = 0.08$).

\subsection{Comparison with M~33}

In a companion paper \citep{gonz12}, we have 
performed a similar analysis on the flocculent galaxy M~33, except that we have
compared the {\em radial} profiles of cluster mass distribution and
gas surface densities. Since M~33 is a flocculent galaxy without a strong
spiral perturbation, in general one can assume that the $\Sigma$ 
azimuthal averages provide a good estimate of the actual surface densities at
any given point located at the same galactocentric distance.    
From the radial analysis, and both the average and median of the 
5 most massive clusters in data bins, we derive for M~33:  
log$_{10}\ M_{\rm mean,3rd} \propto (4.7 \pm 0.4)\ {\rm log}_{10}\ \Sigma_{\rm gas}$;
log$_{10}\ M_{\rm mean,3rd} \propto (1.3 \pm 0.1)\ {\rm log}_{10}\ \Sigma_{\rm H_2}$;
log$_{10}\ M_{\rm mean,3rd} \propto (1.0 \pm 0.1)\ {\rm log}_{10}\ \Sigma_{\rm SFR}$.
One would also derive log$_{10}\ M_{\rm mean,3rd} \propto 
(9.6 \pm 1.2)\ {\rm log}_{10}~\Sigma_{\rm HI}$,
but there is a very small range of $\Sigma_{\rm HI}$ values
in the disk of M~33, as 
the radial distribution of neutral gas has $d~\Sigma_{\rm HI}/dR_{25} 
\approx 0.2 M_\odot {\rm pc}^{-2} R_{25}^{-1}$ 
(exponential scale length $h_{\rm R} = 4.3\ R_{25}$), i.e., 
it is more than twice as shallow as in M~51, which has $h_{\rm R} = 2\ R_{25}$.

\begin{figure*}
\includegraphics[angle=-90.,width=0.85\hsize,clip=]{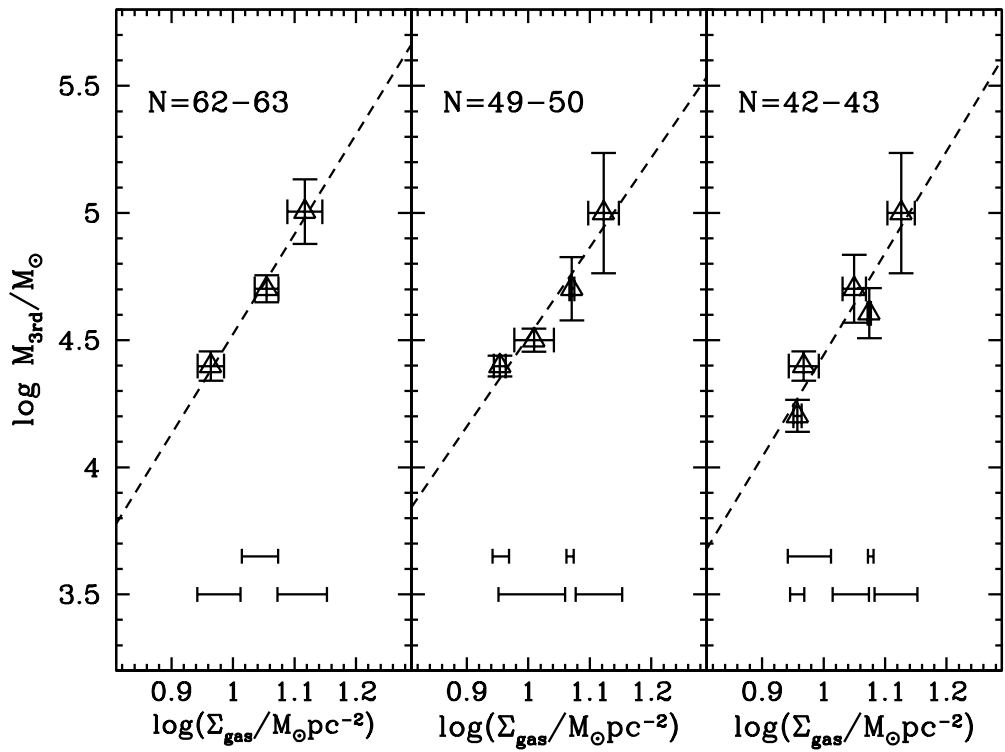}
\caption{M~33, log$_{10}$ mass of third most massive cluster vs.\ log$_{10}$ total gas surface density. 
{\it Empty triangles:} $M_{\rm 3rd}$, median of 5 most massive clusters in bin.
{\it Dashed line:} linear fit. Other symbols as in Figure~\ref{m51_stot}.
}
\label{m33_stot}
\end{figure*}

\begin{figure*}
\includegraphics[angle=-90.,width=0.85\hsize,clip=]{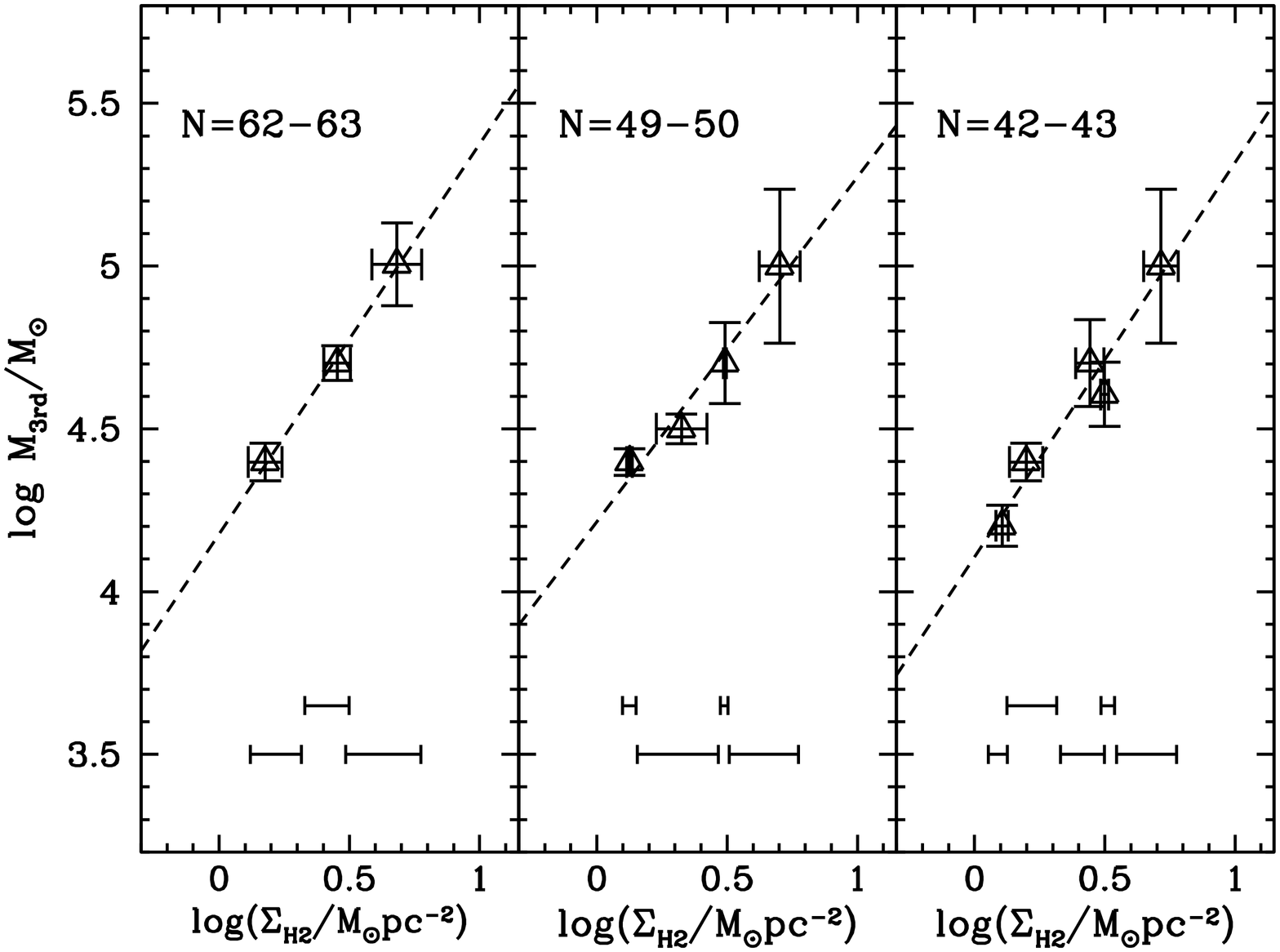}
\caption{M~33, log$_{10}$ mass of third most massive cluster vs.\ log$_{10}$ molecular gas surface density. 
Symbols as in Figure~\ref{m33_stot}.
}
\label{m33_sH2}
\end{figure*}

\begin{figure*}
\includegraphics[angle=-90.,width=0.85\hsize,clip=]{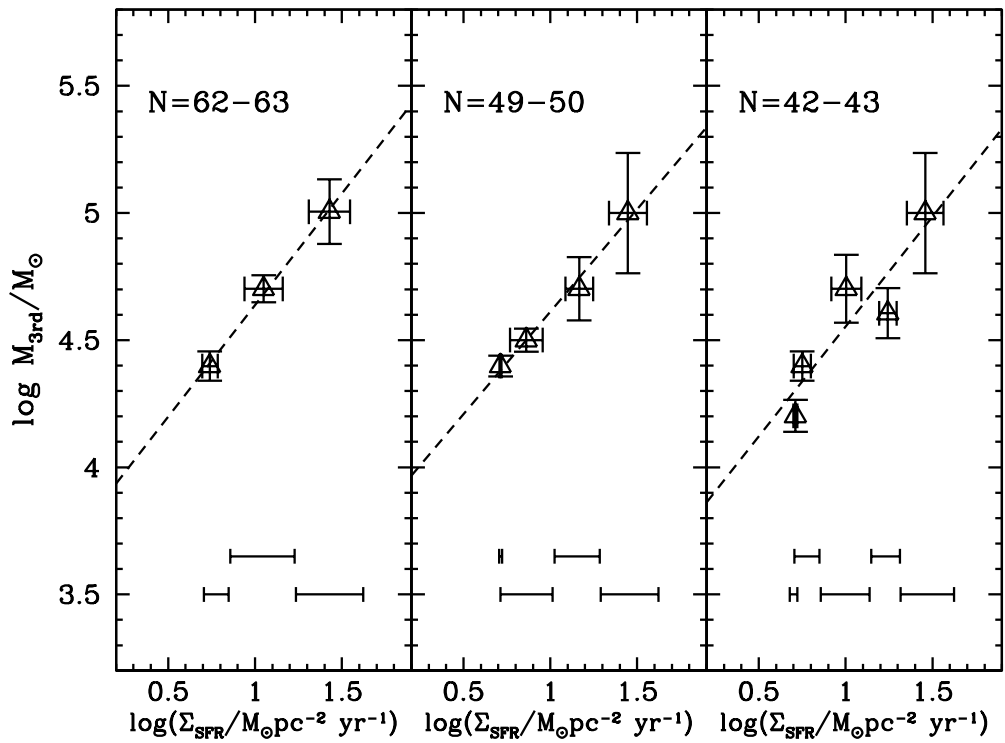}
\caption{M~33, log$_{10}$ mass of third most massive cluster vs.\ log$_{10}$ SFR surface density. 
Symbols as in Figure~\ref{m33_stot}
}
\label{m33_sSFR}
\end{figure*}

\begin{figure*}
\includegraphics[angle=-90.,width=0.85\hsize,clip=]{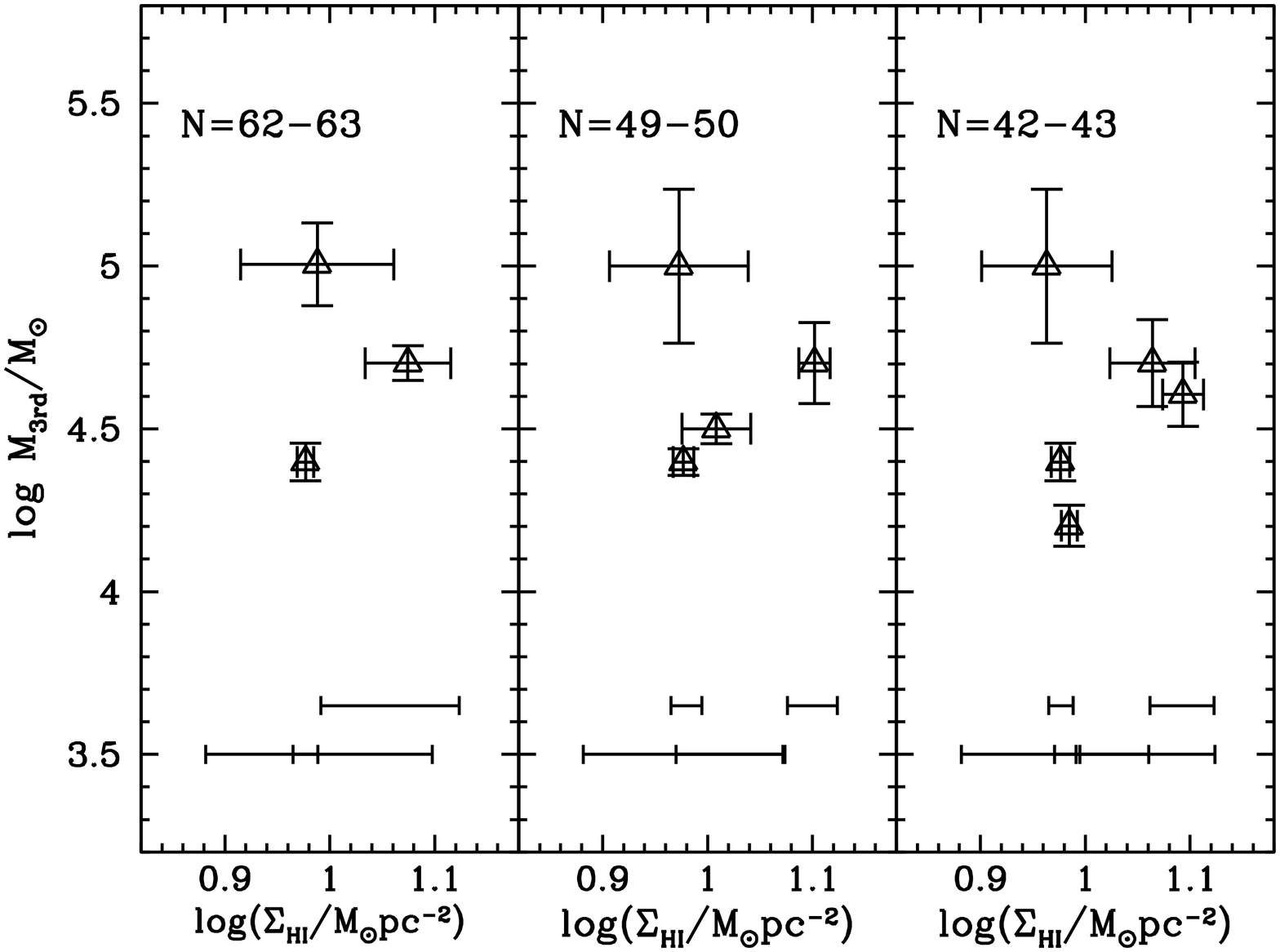}
\caption{M~33, log$_{10}$ mass of third most massive cluster vs.\ log$_{10}$ neutral gas surface density.
Symbols as in Figure~\ref{m33_stot}
}
\label{m33_sHI}
\end{figure*}

Here, we present the M~33 data in a way that is directly comparable
to our analysis of M~51. To this end, and since all ISM 
phases have roughly exponential profiles, we use radius as a proxy
for surface density, and produce plots for M~33 analogous 
to Figures~\ref{m51_stot},~\ref{m51_sH2},~\ref{m51_sSFR}, and~\ref{m51_sHI}. 
We note that 
the log$_{10}~\Sigma$ ranges indicated by the bars at the bottom
of the panels sometimes overlap slightly, owing to the 
departures of the surface density profiles from a perfect
exponential; and (2) we omit in all cases the last bin, i.e.,
the one at the largest galactocentric distance. As explained in
\citet{gonz12}, this bin is
dominated by the most massive cluster in the galaxy,
whose formation conditions are badly misrepresented by 
the azimuthally averaged gas surface density. Moreover,
the last bin always includes a radial range for which
there is no gas data; these stop at $\approx 0.85\ R_{25}$.  
The results of this exercise are shown in 
Figures~\ref{m33_stot},~\ref{m33_sH2},~\ref{m33_sSFR}, and~\ref{m33_sHI}, and in Table 3.
Firstly, this fresh look at the data highlights that 
a correlation does not seem to exist between 
$M_{\rm 3rd}$ and $\Sigma_{\rm HI}$.\footnote{Notice also that
the bars at the bottom of Figure~\ref{m33_sHI} overlap more than
a little, a sign that $\Sigma_{\rm HI}$ is far from changing monotonically
with radius.} As for the other 
ISM components, we get:
log$_{10}\ M_{\rm 3rd} = (3.8 \pm 0.3)\ {\rm log}_{10}\ \Sigma_{\rm gas} + (0.7 \pm 0.3)$;
log$_{10}\ M_{\rm 3rd} = (1.2 \pm 0.1)\ {\rm log}_{10}\ \Sigma_{\rm H_2} + (3.8  \pm 0.1 )$;
log$_{10}\ M_{\rm 3rd} = (0.9 \pm 0.1)\ {\rm log}_{10}\ \Sigma_{\rm SFR} + (4.2 \pm 0.1)$. 

The 2-D correlation of log$_{10}\ M_{\rm 3rd}$ with total gas surface density, log$_{10}\ \Sigma_{\rm gas}$, has a slightly
shallower slope than the comparison between radial profiles. 
The radial and 2-D results for molecular gas and star formation rate densities,
on the other hand, are consistent within 1 $\sigma$. 

\begin{deluxetable}{lrrrrrr}
\tablenum{5}
\tabletypesize{\small}
\tablecolumns{4}
\tablewidth{10cm}
\tablecaption{M~33. Fits to log$_{10}\ (M_{\rm 3rd}/M_\odot)$ vs.\ log$_{10}\ \Sigma$, $\beta^{\prime}_x$ and $\alpha^{\prime}_x$ }
\tablehead{
\colhead{$N_b$}&
\colhead{$N_{\rm cl}$}&
\colhead{$\beta^{\prime}_x$}&
\colhead{$\alpha^{\prime}_x$}\\
}
\startdata
\multicolumn{4}{c}{$\Sigma_{\rm gas}/M_\odot$ pc$^{-2}$} \\ \hline
 3 & 62-63 & $(3.9 \pm 0.4)$ & $(0.6 \pm 0.4)$  \\
 4 & 49-50 & $(3.5 \pm 0.6)$ & $(1.0 \pm 0.6)$ \\
 5 & 42-43 & $(4.0 \pm 0.8)$ & $(0.4 \pm 0.8)$  \\ \hline
\multicolumn{4}{c}{$\Sigma_{\rm H_2}/M_\odot$ pc$^{-2}$} \\ \hline
 3 & 62-63& $(1.2 \pm 0.1)$ & $(4.18  \pm 0.03)$   \\
 4 & 49-50& $(1.1 \pm 0.2)$ & $(4.2 \pm 0.1)$  \\
 5 & 42-43& $(1.2 \pm 0.2)$ & $(4.1 \pm 0.1)$  \\ \hline
\multicolumn{4}{c}{$\Sigma_{\rm SFR}/M_\odot$ pc$^{-2}$ Gyr$^{-1}$} \\ \hline
 3 & 62-63& $(0.9 \pm 0.1)$ & $(3.8  \pm 0.1)$  \\
 4 & 49-50& $(0.8 \pm 0.1)$ & $(3.8 \pm 0.1)$ \\
 5 & 42-43& $(0.9 \pm 0.2)$ & $(3.7 \pm 0.2)$   \\ \hline
\enddata
\tablecomments{Col.\ (1): number of bins. Col.\ (2):
number of clusters in each bin. Col.\ (3): best-fit slope (eq.~1).
Col.\ (4): best-fit intercept (eq.~1). 
}
\end{deluxetable}

\section{Summary and conclusions}

We have analyzed the relationship between maximum cluster mass and gas
surface density in M~51, in order to explore the suggestion that
maximum cluster mass is determined by physical processes, e.g.,
the equilibrium pressure between cluster forming cores and the
ambient interstellar medium \citep{lars02,bill02}, and
since the existence of such a relationship can reconcile SFR measurements
derived, respectively, from H$\alpha$ and FUV emission in galaxy disks. 
Furthermore, we have already found evidence in M~33 that 
star formation is not a stochastic process, and that 
the range of star cluster masses is driven by environmental physics.

In our work, we have used published gas data of M~51 \citep{rots90,schu07,flet11}, and 
catalogs originally comprising more than 1800 young star clusters in its disk, also from the
literature \citep{hwan08,hwan10}. Among other information, these catalogs provide 
positions, ages, reddenings, and masses
for the clusters. 

We have compared the 2-D distribution 
of the young star clusters with those of gas surface density
and star formation rate. 
To test whether any trend is 
caused by random sampling from the cluster mass function as
a size of sample effect, we have measured the
distribution of maximum mass with gas surface density in 2 to 6 bins with an equal number of
clusters in each bin. 

We examined two samples of clusters in M~51: one of 167 clusters 25 to 400 Myr old, and
with a mass of at least 1$\times 10^5 M_\odot$, and another one 
of 56 clusters younger than 10 Myr, with masses of 5000 $M_\odot$ and up.
For the older sample, regardless of the number of bins, 
we find no correlation between the mass of the 3rd most massive cluster in each bin,
$M_{\rm 3rd}$, and the local surface densities of total gas, molecular gas, or star
formation rate. There is, however, a hint of a correlation with 
local neutral gas surface density, such that log$_{10}\ M_{\rm 3rd} =  
(0.4 \pm 0.2)\ {\rm log}_{10}\ \Sigma_{\rm HI} + (5.2 \pm 0.3)$.

For the younger sample, we find no correlation between $M_{\rm 3rd}$, and either
$\Sigma_{\rm H2}$ or $\Sigma_{\rm SFR}$. On the other hand, there is a 
correlation with $\Sigma_{\rm HI}$ that is tighter and with a slightly
steeper slope than the one found for old sample (although still consistent with it): 
log$_{10}\ M_{\rm 3rd} =  
(0.6 \pm 0.1)\ {\rm log}_{10}\ \Sigma_{\rm HI} + (3.5 \pm 0.1)$. 
A correlation may also exist for the most recent burst of $M_{\rm 3rd}$ with
the total mass surface density $\Sigma_{\rm gas}$:
log$_{10}\ M_{\rm 3rd} = (0.5 \pm 0.2)\ {\rm log}_{10}\ \Sigma_{\rm gas} 
+  (3.4 \pm 0.2)$. If true, the comparison between the correlations with
the older and younger burst implies that the normalization depends on
the cluster mass and/or age ranges.
    
The results could not be more different from the ones obtained for the
flocculent galaxy M~33. 
From a 1-D radial analysis, we have previously derived for M~33: 
$M_{\rm 3rd} \propto \Sigma_{\rm gas}^{4.7 \pm 0.4};$
$M_{\rm 3rd} \propto \Sigma_{\rm H_2}^{1.3 \pm 0.1};$
$M_{\rm 3rd} \propto \Sigma_{\rm SFR}^{1.0 \pm 0.1}.$ 
With a treatment that mimicks the 2-D analysis of M~51, we get for M~33:
$M_{\rm 3rd} \propto \Sigma_{\rm gas}^{3.8 \pm 0.3};$
$M_{\rm 3rd} \propto \Sigma_{\rm H_2}^{1.2 \pm 0.1};$
$M_{\rm 3rd} \propto \Sigma_{\rm SFR}^{0.9 \pm 0.1}.$ 

In M~33, the relation between $M_{\rm 3rd}$
and $\Sigma_{\rm SFR}$ is consistent with the expectations from pressure equilibrium
considerations. 
On the other hand, the slope of the correlation with $\Sigma_{\rm gas}$ is 
steeper than the value assumed by \citet{pfla08} to explain the H$\alpha$ cut-off in
galaxy disks. There is no correlation with $\Sigma_{\rm HI}$, 
but M~33 is known to
have been interacting, and the HI distribution
may not reflect the equilibrium configuration.
The tight correlation with $\Sigma_{\rm H_2}$ and $\Sigma_{\rm SFR}$, and the looser one 
with $\Sigma_{\rm gas}$ 
agree with the recent results of a study of
33 nearby spiral galaxies by \citet{schr11}, 
who find a tight and linear relation between $\Sigma_{\rm SFR}$ and 
$\Sigma_{\rm H_2}$, with little dependence on $\Sigma_{\rm gas}$.

How can one understand the diametrically opposite results found for the grand-design spiral M~51?
It is, of course, possible that 
stochastic sampling is at work in M~51, such that the most massive cluster masses scale with the size of
the sample. But then we would have to accept that stochastic sampling operates in
some cases, like in M~51, and not in others, like in M~33.

We propose a different hypothesis.
The large azimuthal variations in gas and
SFR surface densities in M~51 preclude a radial approach.
Instead, we compared star cluster masses 
with local surface densities. The problem then is that, typically, the clusters 
detected in the ACS data with masses above 
the completeness limit of 10$^5 M_\odot$ are older than 10$^8$ yr (only 24 out of 
167 have ages between 40 and 90 Myr). Even if the  
cluster mass function is approximately
independent of age over a timescale of a few hundred million years, clusters also 
complete between half and two full orbits around the galaxy in this time, and hence the 
local gas density around any given cluster probably has little to do with the
gas density at the site and time of cluster formation. 

The two facts, that the correlation between $M_{\rm 3rd}$ and $\Sigma_{\rm HI}$ seems to be tighter for the 
younger sample, and that this latter group of clusters might also exhibit a correlation between    
cluster mass and total gas surface density $\Sigma_{\rm gas}$, give some credibility to this 
hypothesis. Unfortunately, the resolution of the radio data implies that we are comparing
clusters with sizes of the order of 10 pc with gas surface densities averaged over 
an area of $\approx 750 \times 750$ pc$^2$. Even in the case of the younger sample, 
the measured $\Sigma$s are likely quite diluted compared to the actual densities
relevant for the formation of the clusters.

By contrast, 257 out of the 258 clusters kept for the final analysis in M~33 are younger than
2.5$\times 10^7$ yr (the age of the exception is 60 Myr) and, at any rate, 
the lack of a strong spiral perturbation means that 
the azimuthally averaged gas emission 
is in general representative of the star-forming conditions 
of clusters located at the same radii.

To be able to perform a statistically significant analysis of clusters younger
than 25 Myr, still close to their birthsites and whose local gas densities 
resemble their star-forming
environments, the depth and resolution of the cluster data need to 
be enough for the assembly of a sufficiently numerous sample above the mass function
completeness limit.\footnote{Number 
of clusters increases
with decreasing mass. On the other hand, the absolute number of the most massive, 
easily detectable, clusters will grow 
as generations of star formation accumulate. Hence, the majority
of massive clusters in a galaxy, even with active star formation,
will be old, whereas for this project it is best to 
work only with the most recent bursts.} The gas data should also have
a resolution comparable to the physical sizes of the clusters.

The confirmation of the M~33 results will have to wait
for the outcome of studies in closer targets, like the Magellanic Clouds, where
the mass density distribution of clusters younger than a few $\times 10^7$ yr can be probed
with large numbers down to 100-1000 $M_\odot$, and where the clusters' physical sizes are
well matched by the resolution
achievable at radio wavelengths; the data to do so already exist.

\acknowledgments
RAGL is grateful for the support received from DGAPA, UNAM,
and CONACyT, Mexico.
N.\ Hwang and M.\ G.\ Lee generously provided the cluster catalogs 
used in this work.
We thank A.\ Rots, K.\ Schuster, and R.\ Beck for
granting access to the HI, CO, and 20 cm continuum images of M~51, respectively.
L.\ Loinard helped with the nitty gritty details of the conversion
of the reduced radio data to physical surface densities, and with
the transformation of the HI image to coordinates J2000.
We acknowledge the positive and helpful comments of an anonymous referee.

\end{document}